# Configuration Interaction of Hydropathic Waves Enables Ubiquitin Functionality

Douglas C. Allan[1] and J. C. Phillips[2]

1. Corning Inc, Div. Sci. & Technol., Corning, NY 14831 USA
2. Dept. of Physics and Astronomy, Rutgers Univ., Piscataway, N. J., 08854

Abstract

Ubiquitin, discovered less than 50 years ago, tags thousands of diseased proteins for destruction. It is small (only 76 amino acids), and is found unchanged in mammals, birds, fish and even worms. Key features of its functionality are identified here using critical point thermodynamic scaling theory. These include Fano interference between first- and second-order elements of globular surface shape transitions. Comparison with its closest relative, 76 amino acid Nedd8, shows that the latter lacks these features. A cracked elastic network model is proposed for the common target shared by many diseased proteins.

Between the 1960s and 1980s, most life scientists focused their attention on protein creation by DNA and the translation of its coded information. Protein degradation was a neglected area, considered to be a nonspecific, dead-end process. Although it was known that proteins do turn over, the large extent and high specificity of the process, whereby distinct proteins have half-lives that range from a few minutes to several days, was not appreciated. The discovery of the complex cascade of the ubiquitin pathway revolutionized the field [1]. Today attention continues to be focused on DNA-based personalized medicine, but the very size of DNA makes this a challenging prospect.

The extent and especially the depth of our understanding of individual proteins continue to grow in unexpected ways. Ubiquitin, discovered as the protein that targets other proteins for degradation [2], is the subject of multiple international conferences annually. Here we apply theoretical tools that show that ubiquitin has something very special in combination with hemoglobin: it confirms a remarkably prescient conjecture by Hopfield in 1973 [3] that identified a fundamental feature that makes both proteins necessary metabolic constituents of all tissues. He suggested that "a quantitative understanding of co-operativity in hemoglobin must include a description of where and how the free energy of co-operation is stored in the molecule. One extreme possibility is that the free energy is stored as small amounts of strain energy in many bonds, so that all bonds are almost normal." His "linear distributed energy model" is consistent with Fano's configuration interaction model of atomic spectra [4]. Specifically, it supports the existence of, and interference between, strain waves localized at hemes with continuum strain waves associated with tetrameric globin interfaces in hemoglobin [5].

Here we will continue to use thermodynamic scaling and hydropathic waves [6] to analyze the functionality of ubiquitin. Whereas hemoglobin is a ~ 580 amino acid tetramer - a dimer of dimers of ~ 145 amino acid globins - ubiquitin is a small protein with only 76 amino acids. Given the weakly broken tetrahedral symmetry of hemoglobin, one could guess that the globins are coupled by hydropathic waves, so the linear [3] interference model is plausible, especially after it has been supported by parameter-free calculations [5]. Because ubiquitin is so small, why should we expect it to exhibit wave interference? The answer is that ubiquitin's ability to identify diseased proteins suggests that these protein networks share a common topological feature. Their strain fields are cracked, and interference between distributed continuum waves and waves localized at nanocracks [7] could be a second example of the Fano effect [4].

Because they are self-organized, glass and protein networks share many thermodynamical properties [8-12]. Their structural phase transitions are generally mixtures of first-order and second-order transitions, and can have predominantly first-order or second-order character. In glasses this distinction has been used to study dynamical relaxation [13]. In proteins the same distinction can be made by using two hydropathicity scales, for instance the classical scale associated with first-order water-air unfolding enthalpy differences [14], and the modern second order scale associated with fractals derived from log-log plots of solvent associated surface areas as a function of segmental length [15]. Often the modern scale is more accurate, but exceptions may occur with enzyme interactions.

Another tool which has provided some striking results is the evolution of a protein, for instance, Hen Egg White [6]. However, ubiquitin is very special, because its 76 amino acid sequence has been found unchanged in mammals, birds, fish, and even invertebrate worms (search on Web-based BLAST, the Basic Linear Alignment Search Tool). Absent evolutionary guides, one still has length as a probe for long-range interactions [13,16,17]. Protein interactions are usually strong at hydrophobic spots, so the length scale of globular surface roughness (in other words, the variance of hydropathic profiles) is a key tool for quantifying hydropathic waves and their interference.

All calculations are based on the hydropathic values Ψ(amino acid) [14, 15] linearly scaled to a common center and a common range [18] for each of the 20 amino acids. These are then converted to a triangular matrix Ψ(aa,W), where W is the length of a sliding window centered on each protein site. Variance is a standard statistical tool, available on EXCEL, where all calculations can be performed. We have studied the full range $1 \leq W \leq 75$.

Absence of evolution is the most striking feature of ubiquitin; thermodynamically it is explained by saying that ubiquitin reached its critical point of functional perfection already even in worms. This conclusion can be tested by comparing the profiles of ubiquitin and Nedd8 (its closest sequence-alike (~ 70%) protein [1,2]). Nedd8 (76 amino acids) itself is interesting: it evolved very little from worm to mouse (BLAST sequence similarity high, ~ 92%), and remained unchanged in mammals from mouse to humans. The perfection of ubiquitin (76 amino acids) is reflected in the ~ 80 biological processes listed on its Uniprot page, compared to 9 for Nedd8, and 5 for Hen Egg White (148 amino acids, almost twice as large).

The broad features of the Ψ(aa,W) ubiquitin profiles with the first-order KD and second-order MZ scale are similar, as shown in Fig. 1, which explains their orientation. This figure shows that the key role in ubiquitin is played by the hydrophobic peak, which functions mechanically as an elastic pivot in the center, separating two hydrophilic hinges. The significance of the hydrophobic peak is also clear in a hydrodynamic model, where peptide chain segment $C_\alpha$ center motion is described by vectors describing particle flow. A general theorem of vector fields decomposes them into solenoidal and irrotational components. The solenoidal components are connected to vortex centers, and here the hydrophobic peak is such a center. Its relative strength is larger using the first-order KD scale.

The ratios of the globular roughnesses with the two scales are shown in Fig. 2, whose caption discusses their structural details. Here one should note that globular roughness is a property currently fashionable in pure mathematics (differential geometry). It is related to volume properties by a generalized version of Stokes' theorem, a standard part of advanced calculus. Spectroscopic data were the basis for Niels Bohr's planetary model of the atom, but it was not until 50 years later that it was realized that bilinear opto-electronic wave interference can be found in many carefully prepared liquid and solid samples [4].

Most proteins' functionality is complex and extremely nonlinear, so one would not expect to find bilinear wave interference in their properties, especially in hydropathically shaped globular structures. Hemoglobin and ubiquitin are exceptions, the first identified (to our knowledge). The long-standing mystery of the molecular origin of collective heme α - β interactions is explained by interference of heme and interface-based waves [5]. Here the perfection and functional generality of tagging strain field cracks of diseased proteins has led to the rich structure of ubiquitin in Fig. 2, which is basically absent in Nedd8.

As a footnote, there must be some indications of how the magical structure of ubiquitin has evolved, hidden in Nedd8 properties. There is, because although ubiquitin is perfect and has not evolved, its look-alike Nedd8 has. It turns out that the largest worm-human evolutionary changes in the $\Psi(aa, W = 9)$ Nedd8 profiles occur in the first-order KD profile, which is shown in Fig. 3. The variance profiles of human and worm Nedd8 are compared in Fig. 4 and discussed there.

With respect to the universality of ubiquitin tagging, cracks in a protein network can cause protein dysfunction, while preserving the protein fold. One can suppose that the cracks involve packing misfits, which would enhance a few irreversible first-order (strong) interactions, compared to many more reversible second-order (weak) allosteric interactions. Cracks can occur between rigid homologous domains, and may involve shear [19-21]. At such cracks thermal vibrational amplitudes are small, or alternatively there is less thermal noise, facilitating long-range (large W) allosteric binding of ubiquitin along a crack. The crack length can be estimated to be about half of a typical domain length in giant hub enzymes such as tyrosine kinases [22]. The latter domain lengths L are around 100 amino acids; halves of these lengths fit the resonances in Fig. 2. The lengths W of the localized modes associated with the cracks would also tend to be around L/2 for the softest modes. These same modes would interfere with conformational motions associated with protein functionality. Cracking leads to a counterintuitive catalytic effect of added denaturant on allosteric enzyme function in adenylate kinase [19,20].

In conclusion, the notion that wave interference effects can be observed in connection with strain fields as well as in opto-electronic fields has become apparent from rapid advances in nanoscience [7]. Even so, the observation of such delicate effects in connection with hydropathic waves requires choosing just the right proteins as examples, to be combined with just the right bioinformatic scaling tools [15]. Although such ideal examples of wave interference effects are few, they form a benchmark against which many other concepts and theoretical tools can be tested. The present approach to ubiquitylation, which emphasizes long-wave length attractive interactions mediated by hydropathic waves, should be contrasted with contact interaction models based on static structures [23]. Perhaps the combination of the two approaches will yield new insights.

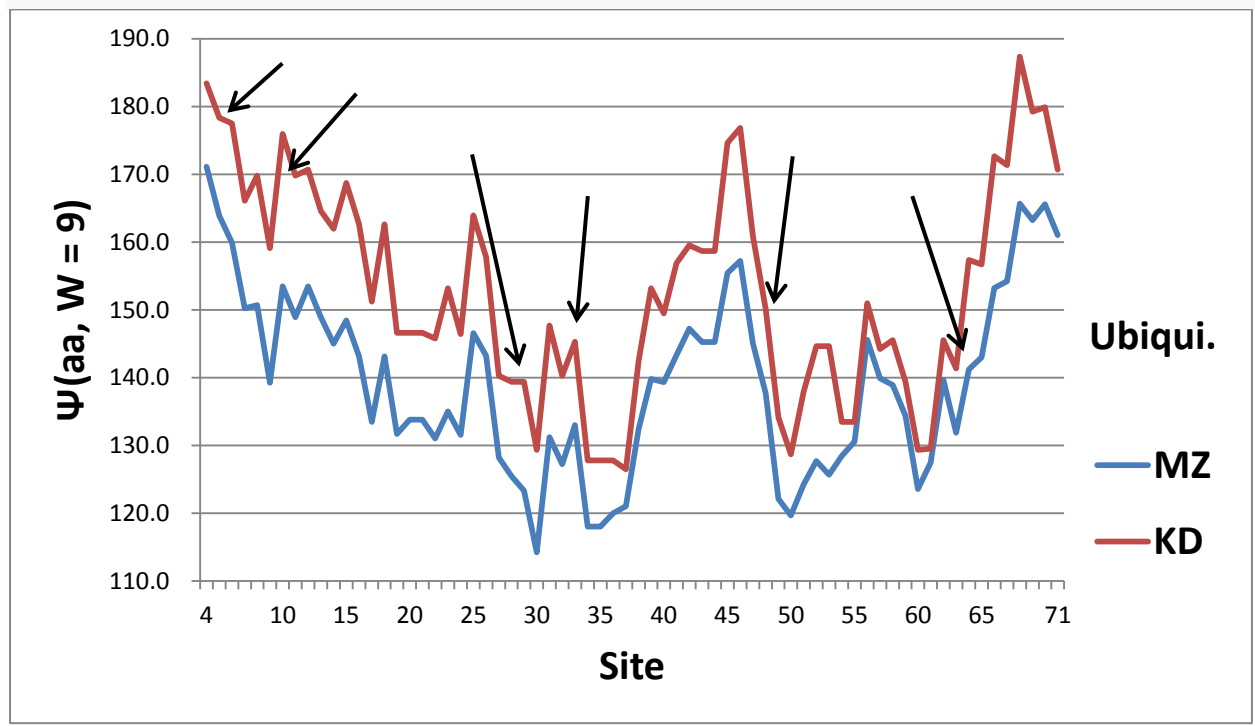

Fig. 1. Hydropathic profiles with sliding window width W = 9 (smallest value associated with MZ scale, and highest resolution) for ubiquitin. Sites are numbered from N terminal to C terminal, and the two scale centers have been slightly offset for clarity, with larger values being more hydrophobic, and smaller values more hydrophilic. Hydroneutral is approximately 155 on both scales The overall shape is roughly that of a parabolic bowl, with a hydrophobic peak ~ 45 functioning as an elastic pivot in the center. Thus the N and C terminal wings can swing separately, facilitating tagging the opposite sides of a crack in the strain field of a diseased protein. The arrows mark lysine (K) sites, which have characteristically specific covalent lysine targets. Lysine is the most hydrophilic amino acid on almost all scales, and here its various roles are determined by target crack hydrophilic properties. The average hydropathicity of ubiquitin is hydrophilic, which means it is elastically softer than average, which again facilitates tagging diseased cracks.

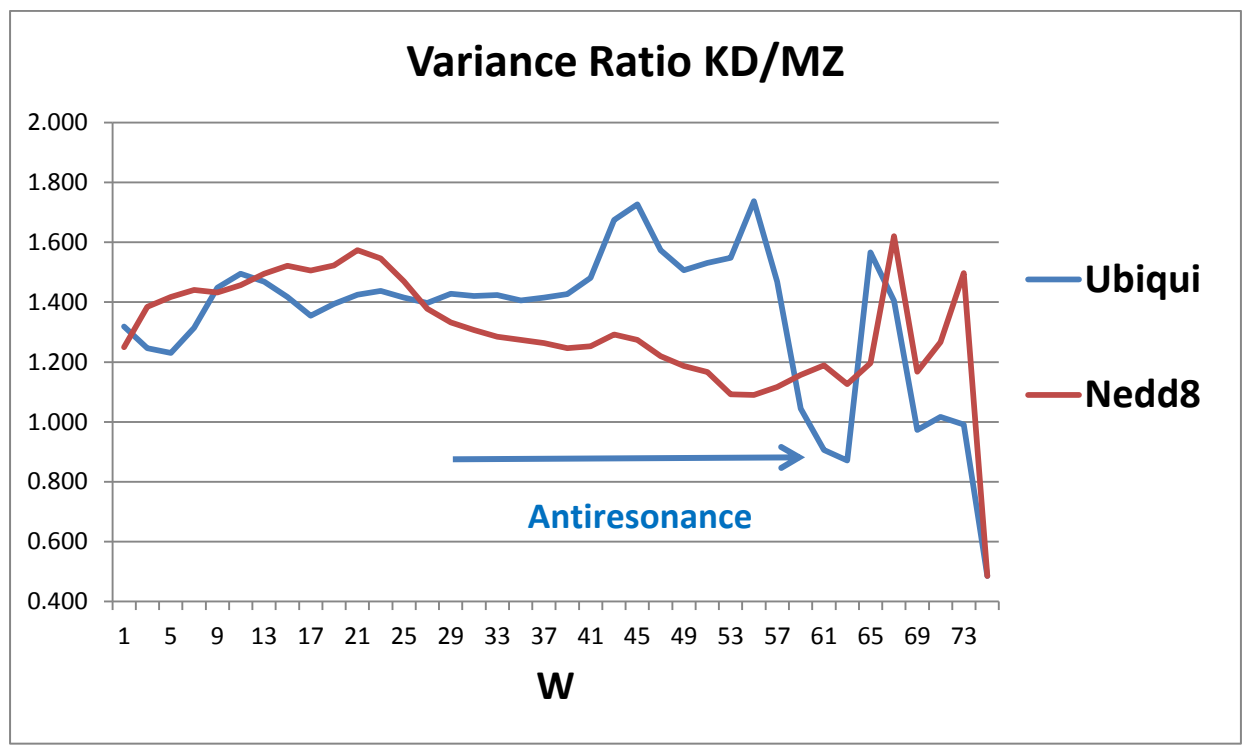

Fig. 2. The human globular roughness, as a function of sliding window length W, shows rich structure for ubiquitin, but only weak structure for Nedd8. Here the structure above W 67 could be noise, but the two peaks centered on W = 45 and 55, and the sharp dip at 61-63, are genuine. Because ubiquitin is perfect, we expect that its second-order fine structure, monitored by the MZ variance, has become smooth (small variance), and that it is this MZ smoothness being small (or the KD roughness being large) that produces the two peaks in KD/MZ. Similarly, the 61-63 antiresonance should be caused by a rough spot in MZ which evolution has not removed. This could be the two N- and C-wings of the overall profile (Fig. 1) which are critical to its tagging opposing sides of target cracks.

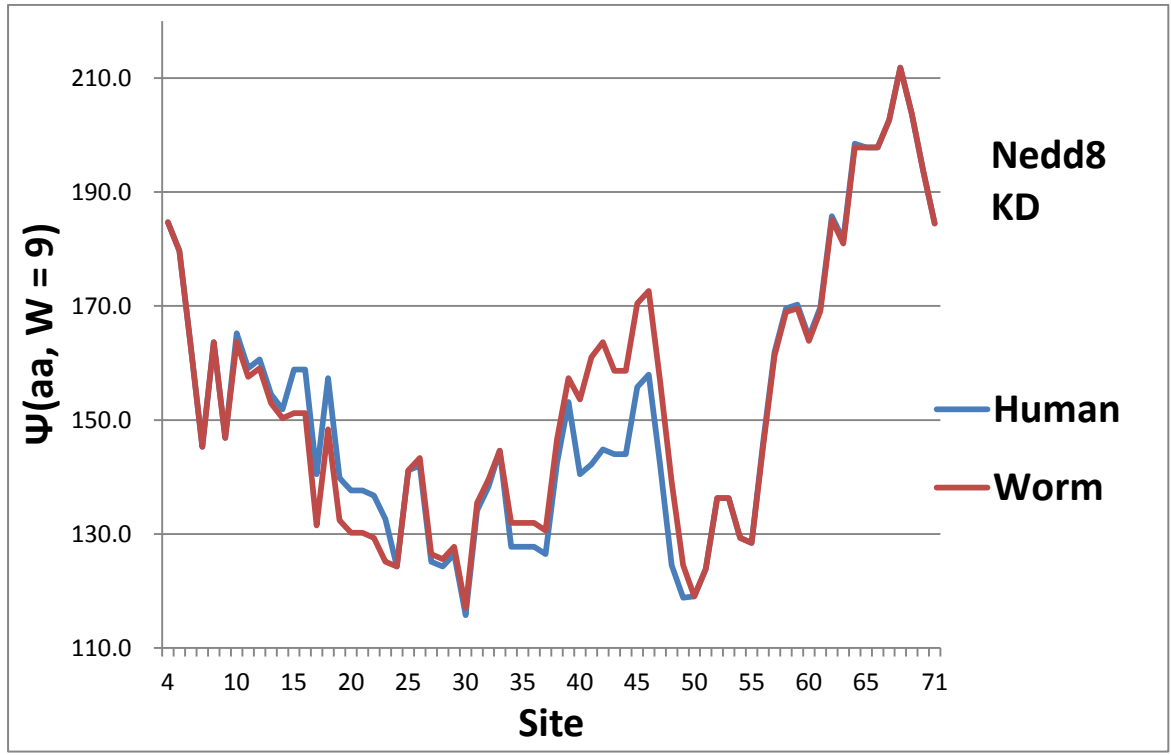

Fig. 3. In the evolution of Nedd8 from worm to human, most of the changes appear to be first order (KD scale) and are concentrated in softening the central pivot. This contrasts strongly with distributed smoothing of ubiquitin in Fig. 1.

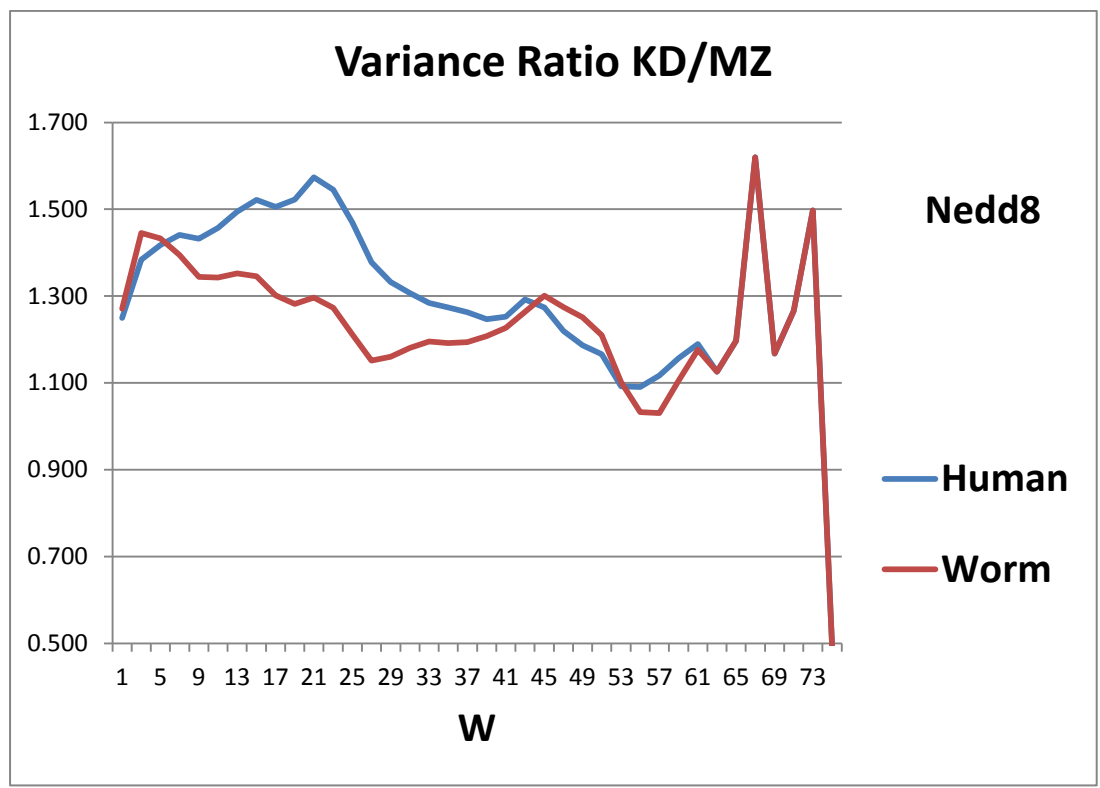

Fig. 4. Comparison of the roughness (variance) of the hydropathic profiles for different sliding window wave lengths W for worm and human Nedd8 shows that evolution has mainly improved the shorter lengths W ~ 20, while the peaks and antiresonance seen in Fig. 2 for ubiquitin are undeveloped.